# Horizontal Layer Constrained Attention Neural Network for Semblance Velocity Picking

Chenyu Qiu[①], Bangyu Wu[①]*, Meng Li[②], Hui Yang[③] and Xu Zhu[①]

[①] School of Mathematics and Statistics, Xi'an Jiaotong University, Xi'an, 710049, China

[②] PetroChina Research Institute of Petroleum Exploration & Development, Beijing, 100083, China

[③] Department of Earth and Space Sciences, Southern University of Science and Technology, Shenzhen, 518055, China

* Corresponding author: *Bangyu Wu (bangyuwu@xjtu.edu.cn)*

*Abstract*—Semblance velocity analysis is a crucial step in seismic data processing. To avoid the huge time-cost when performed manually, some deep learning methods are proposed for automatic semblance velocity picking. However, the application of existing deep learning methods is still restricted by the shortage of labels in practice. In this paper, we propose an attention neural network combined with a point-to-point regression velocity picking strategy to mitigate this problem. In our method, semblance patch and velocity value are served as network input and output, respectively. In this way, global and local features hidden in semblance patch can be effectively extracted by attention neural network. A down-sampling strategy based on horizontal layer extraction is also designed to improve the picking efficiency in prediction process. Tests on synthetic and field datasets demonstrate that the proposed method can produce reasonable results and maintain global velocity trend consistent with labels. Besides, robustness against random noise is also tested on the field data.

## I. INTRODUCTION

SEMBLANCE velocity analysis is a crucial step in seismic data processing [1]. By picking the center of energy heaps in semblance generated from seismic data, velocity model for normal moveout (NMO) correction can be obtained [2]. When performing manually, it is labor intensive with results susceptible to the experience of experts. In recent years, many automatic semblance velocity picking methods have been proposed to reduce the huge cost of manpower. Fish and Kusuma [3] employ the non-convolutional neural network to choose correct semblance cells for picking. Wei et al. [4] utilize the k-means algorithm which can find and map the clusters on semblance to time-velocity (t-v) pairs. Waheed et al. [5] compare the Density-Based Spatial Clustering of Applications with Noise (DBSCAN) with k-means method in [4] and show the robustness of DBSCAN for automatic velocity picking. Wang et al. [6] propose an adaptive thresholding constraint strategy combined with k-means to obtain a smooth velocity profile. However, these clustering-based methods usually require tedious selection of parameters and are difficult to maintain robustness and stability on complex dataset.

Inspired by the high efficiency and robustness of deep learning in the field of image processing [7]-[8], some methods with different learning strategies and network architectures have been proposed for semblance velocity picking. Park and Sacchi [9] take semblance velocity picking as an image classification problem and use convolutional neural network (CNN) to classify the energy heaps corresponding to a specific velocity. Wang et al. [10] verify both classification and regression strategy in U-Net and conclude that regression achieved higher accuracy for velocity picking. Zhang et al. [11]

utilize You-Only-Look-Once (YOLO) to detect the energy heaps and incorporate the Long-Short Term Memory (LSTM) to obtain the velocity curve. Based on the idea of [11], Ma et al. [12] propose the fit line detection-point adjust (FLD-PA) method to obtain the velocity curves by object detection and adjust the velocity according to the spatial correlation of adjacent semblance. Araya-Polo et al. [13] employ a 3D semblance cube as the input of CNN to reconstruct a 2D velocity profile. Above deep learning methods have made great progress in automatic semblance velocity picking. However, these achievements have relied on the fact that optimization of these deep, high-capacity models requires many iterative updates with sufficient labeled semblance [14]. It makes deep learning fall into a bottleneck in practical application for semblance velocity picking.

In this paper, we propose an attention neural network combined with a point-to-point regression strategy to achieve automatic semblance velocity picking. Through predicting semblance patch in time window to a velocity, limited labeled time-velocity pairs can be fully utilized and easily combined with data augmentation strategy for the training process. Meanwhile, a down-sampling strategy based on horizontal layer extraction is also developed to reduce the number of prediction points. We apply the horizontal layer constrained attention neural network to both synthetic and field data to demonstrate the picking performance.

## II. METHODOLOGY

### A. Point-to-Point Regression Strategy

For semblance velocity picking, most deep learning methods directly employ complete semblance as network input. However, the field data only includes several dozen labeled semblances which means the number of training set is usually not sufficient. Conventional learning strategy breaks down in the small data regime when it tries to learn from very few labeled data. Generally, a semblance will be manually marked as several (t-v pairs. Referring to manual velocity picking, we employ a patch of semblance as input. The patch center is corresponding to the time point for velocity picking. Due to high accuracy using regression strategy [10], NMO velocity value is network output. Consequently, the process of picking t-v pair can be taken as a point-to-point (patch center time point to velocity point) regression. We adopt the Mean Squared Error (MSE) loss as the loss function, can be written as:

$$\boldsymbol{Loss} = ||\boldsymbol{y} - \hat{\boldsymbol{y}}||_2^2 \quad (1)$$

where $y$ and $\hat{y}$ are predicted and labeled velocity, respectively. In this learning strategy, the number of training set is proportional to the number of time samples which is beneficial to enrich the features and amount of training set. And data augmentation strategies such as interpolation, semblance patch reversal [15] can be easily implemented. Taking the relation between semblance and reference velocity which is used to produce semblance into consideration, we also normalize the semblance patch to same scale with reference velocity. The function is expressed as:

$$S_n = \frac{(S - S_{max}) \times (V_{max} - V_{min})}{S_{max} - S_{min}} + V_{min} \quad (2)$$

where $S_n$ and $S$ are the normalized and original semblance patch, respectively; $V$ is the reference velocity for the semblance patch.

### B. Attention Neural Network

Compared with complete semblance, local semblance patch loses some global feature in time dimension though it can bring some advantage in making full use of limited labels. To capture remaining global feature hidden in patches of semblance, a CNN including the attention module is designed. The main modules of network are shown in Fig. 1.

First, the pre-extracted layer (Fig. 1a) is employed to increase the number of channels for multiscale feature extraction from the designed input. The attention module (Fig. 1b) allows CNN to place high weight on regions highly related to the task and do the inference accordingly [16]. We then insert this module into CNN for explicitly modeling long range dependencies and fully capturing global feature. In order to extract high-level feature and gain accuracy from increased depth of network [17], the residual modules (Fig. 1c) are connected behind. Finally, average pool layer and full connection (FC) layer are added to resize the network output. The architecture of attention neural network is shown in Fig. 2.

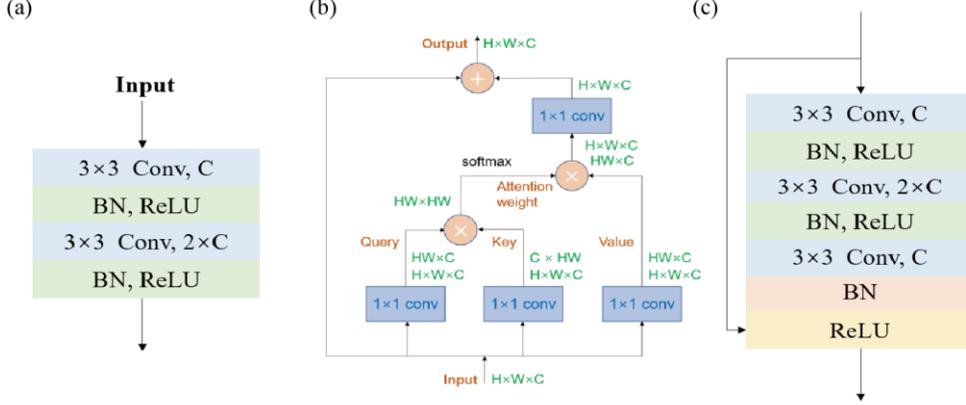

Fig. 1. Main modules in proposed velocity picking CNN. (a) Feature pre-extraction module; (b)Attention module; (c) Residual module.

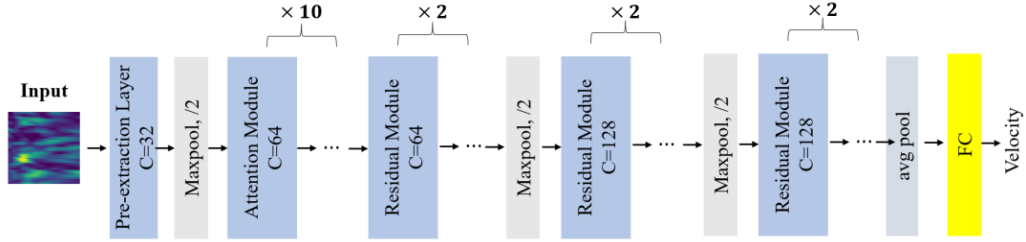

Fig. 2. The attention neural network architecture for automatic velocity picking from semblance.

## C. Down-sampling Strategy Based on Horizontal Layer Extraction

The attention neural network can map a semblance patch into NMO velocity value. The velocity curve can be predicted point by point which cost the extensive calculation. To reduce the number of prediction points, we develop a strategy based on horizontal layer extraction down-sampling (HLEDS) utilizing the layer feature hidden in the 3D semblance cube. In this way, attention neural network only needs to predict velocities at several points corresponding to latent horizontal layers. The process of horizontal layer extraction is described below.

Assuming that a dataset including $K$ Common Middle Point (CMP) gathers. The corresponding semblance matrix $S_k$ is defined as follows:

$$S_k = \begin{pmatrix} S_{k,11} & \cdots & S_{k,1m} \\ \vdots & \ddots & \vdots \\ S_{k,n1} & \cdots & S_{k,nm} \end{pmatrix} \quad (3)$$

where $n$ is the number of sampling points in $S_k$, $m$ is the width of $S_k$ and $k$ is corresponding to the $k_{th}$ semblance. Sliding the time window $w(i)$ on $S_k$, an energy spectrum $H_{k,i}$ can be obtained as following:

$$H_{k,i} = S_k \cdot w(i) \tag{4}$$

where $w(i)$ is the window function over time $i$. The $w(i)$ is represented as:

$$w(i) = \begin{cases} 0, & i - \frac{d_w}{2} < i < i + \frac{d_w}{2} \\ 1, & others \end{cases} \tag{5}$$

where $d_w$ is the length of sliding window. $I_k$ is then calculated by marking the time index of the maximum value from windows at each time. Then $I_k$ is represented as:

$$I_k = \{i | H_{k,i}(i,j) = max(H_{k,i}), i \le n\} \tag{6}$$

where $i\ and\ j$ are the abscissa and ordinate, respectively. Repeating the process on all semblances, the section of horizontal layers $L$ with dimension $(n, K)$ can be constructed. L is defined as:

$$L(i,k) = \begin{cases} 1, & i \in I_k \\ 0, & others \end{cases} \tag{7}$$

Applying the Gaussian operator $G$ to smooth $L$ first and then using the upper 75 quantiles to filter the noise out, the result $L^*$ can be obtained as:

$$L^* = filter(L \odot G) \tag{8}$$

where $filter(\cdot)$ sets the value of $L$ less than the upper 75 quantiles to 0 and $\odot$ is the smoothing operator. Select the abscissa and ordinate of the points not equal to 0 in $L^*$ as the candidate point set $X$ which is defined as following:

$$X = \{(i,j) | L^*(i,j) \neq 0, i \le n, j \le K\} \tag{9}$$

Use DBSCAN to eliminate the discrete point in $X$ and obtain the $X'$ which is defined as following:

$$X' = \{(i,j) | (i,j) \in X\ \&\ (i,j) \notin Noise\} \tag{10}$$

Mark the points in $X'$ on the section with values set to 0. The updated $L$ can be written as:

$$L(i,k) = \begin{cases} 1, & (i,j) \in X' \\ 0, & others \end{cases} \tag{11}$$

After the dilatation process in the horizontal dimension, the points in $L$ that continuously exist in the vertical dimension are refined and achieve the final section of horizontal layer $\overline{L}$. These steps are concluded as follows:

Step 1: $Initialize\ \overline{L} = 0, i = 1\ and\ j = 1, where\ i \in [0, n]\ and\ j \in [0, K]$.
Step 2: $i = i + 1\ until\ i = n. Then\ set\ i = 0\ and\ j = j + 1$.
Step 3: $if\ \exists i_a \sim i_b\ s.t. L(i,j) = 1\ and\ a \sim b\ are\ continious,$ set $\overline{L}(\frac{i_a+i_b}{2}, j) = 1$.
Step 4: $if\ i = n\ and\ j = K, output\ \overline{L}$.

### III. EXPERIMENTAL RESULTS

In this section, we employ the attention neural network (Fig. 2) to the synthetic and field data. All experiments carried out the following training setting. The batch size is set to 32. The number of epochs is set to 50. For the first 20 epochs, the learning rate is set to 0.001 and will be halved every 10 epochs afterwards. To avoid overfitting, the network training will be stopped when the validation loss starts to rise. In addition, Adam optimizer [18] is used to optimize the parameters in the network and the parameter weight decay is set to $10^{-7}$. A single GTX2080Ti is used for the network training. The data processing including CMP sorting, mute, NMO and stack are all handled using the Seismic Unix.

#### A. Synthetic Dataset

The benchmark data from SEG Hess model consisting of 720 shot gathers with 656 receivers per-shot generated using the finite-difference modeling software developed at SEP, Stanford University. Among them, 501 CMP gathers are generated and sorted at mid-point positions from (x, z) = (30000,0) ft to (40000,0) ft with offsets from 0 ft to 26200 ft. The time sample interval is 6 ms with 600 samples for each trace. These CMP gathers are then transformed into semblance. Totally 126 semblances at equal CMP intervals are selected for manual velocity picking. The number of manual (t-v) picks is 1168.

Fig. 3 shows two semblances corresponding to CMP 32000 (Fig. 3(a)) and CMP 38000 (Fig. 3(b)) with auto-picks, and manual picks superimposed. It can be seen clearly that the trends of auto-picks are consistent with the manual picks and the picks extracted by HLEDS almost cover the manual picks. Meanwhile, the auto-picks can effectively avoid the energy heaps caused by multiples (around 1.4s and 1.7s in Fig. 3(a) and (b)). Fig. 4 (a) and (b) show the velocity models without and with HLEDS, respectively. The velocity model generated by the interpolation of manual picks is shown in Fig. 4(c). As shown Fig. 4, the velocity models consisting of auto-picks present less lateral and vertical volatility than labels and velocity model with HLEDS performs best in global continuity.

Fig. 5 mainly shows the stack sections and two CMP gathers (CMP 32000, CMP 38000). Fig. 5(a) presents the latent horizontal layers extracted by HLEDS and two CMP gathers before NMO correction. The stack sections (left) and CMP gathers after NMO correction (right) in Fig. 5 (b)-(d) are generated from the velocity models in Fig. 4 (a)-(c). Both Fig. 5 (b) and (c) show the good performance on NMO correction and stack. Especially, the stack sections in Fig. 5 (b) and (c) present continuous and clearer seismic events than labels. Comparing the horizontal layers in Fig. 5 (a) with stack sections, most horizontal layers on stack sections are correctly extracted.

B.   Field Dataset

The second example shows the application of the proposed method on real marine seismic data. A 2D inline section is selected for this experiment. There are totally 1400 shots with 240 receivers each on this inline towing from left to right, with a source and receiver interval of 25 meter, respectively. The time sampling interval is 2 ms with a recording length of up to 6 seconds. Totally 17 semblances at equal CMP intervals are selected for manual velocity picking. The shot gathers can be sorted and filtered into 801 CMP gathers with number of traces more than 60. The number of manual (t-v) picks is 201.

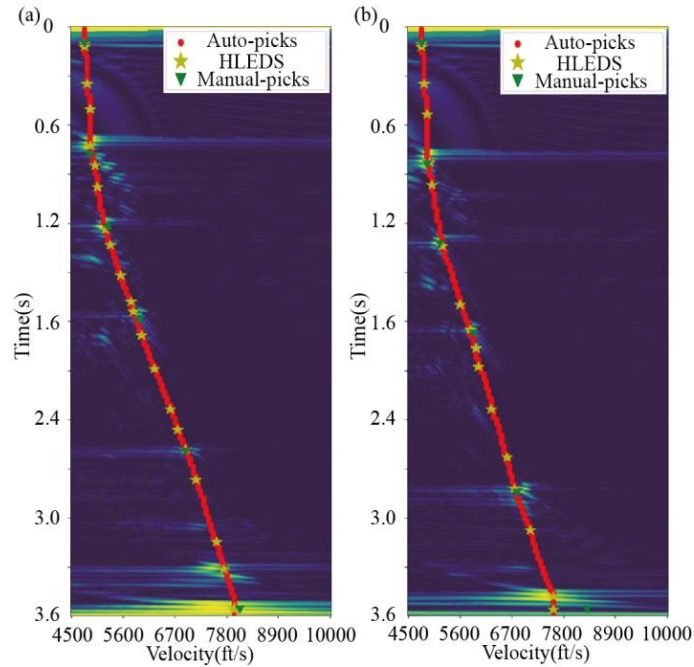

Fig. 3. Comparison among auto-picks without HLEDS, with HLEDS and manual picks on semblances. (a) CMP 32000. (b) CMP 38000.

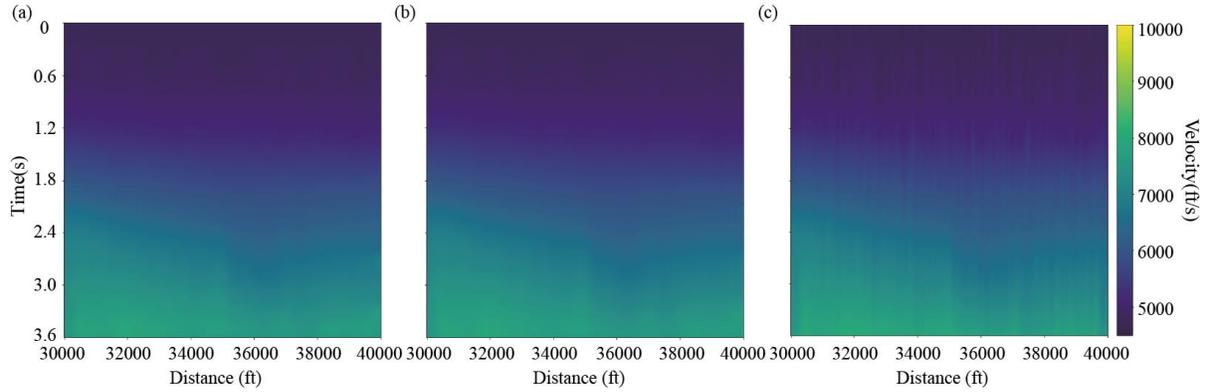

Fig. 4. Velocity models picking from semblances. (a) auto-picks without HLEDS. (b) auto-picks with HLEDS. (c) Label.

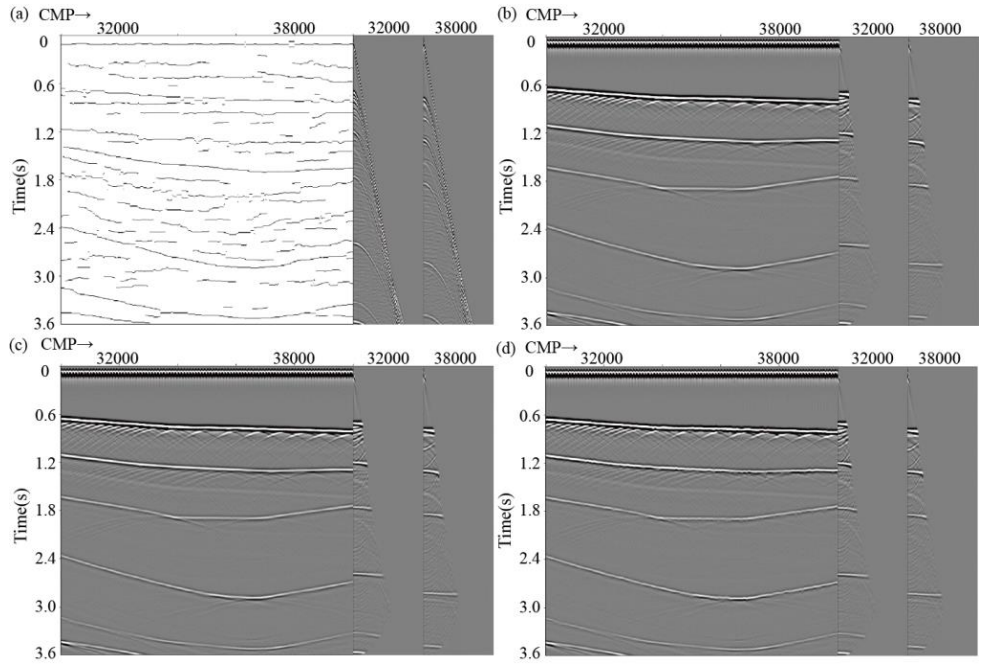

Fig. 5. (a) Latent horizontal layers extracted by HLEDS (left) and two CMP gathers before NMO correction (right). Figures in (b), (c) and (d) are the stack sections (left) and two CMP gathers after NMO correction (right) from auto-picks without HLEDS (b), auto-picks with HLEDS (c) and Label (d), respectively.

## C. Robustness of proposed method against random noise

We also test the robustness of the proposed method against random noise. We add eight different levels of Gaussian noise with signal to noise ratio (SNR) of -10, -5, 0, 5, 10, 15, 20 and 25 dB into the CMP 1179 in field dataset and obtain the prediction results by the proposed method. The NMO velocity curves of the predictions in CMP 1179 with different SNR are shown in Fig. 9. It can be seen that the accuracy and stability of the proposed method decreases as the SNR decreasing from 25 to -10dB. However, the proposed method obtains high accuracy in shallow layer and maintains appropriate trends of velocity curve. Besides, the HLEDS added in Fig. 9 (b) obtains velocity curves smoother. In short, the proposed method shows well robustness against random noise.

TABLE I
COMPARISON OF TIME COST IN PREDICTION WITH AND WITHOUT HLEDS

| Processing Dataset | Prediction without HLEDS | Prediction with HLEDS | |
|---|---|---|---|
| | | HLEDS | Prediction |
| Synthetic datasets | 230.6s | 1.7s | 6.2s |
| Field datasets | 432.8s | 2.3s | 10.1s |

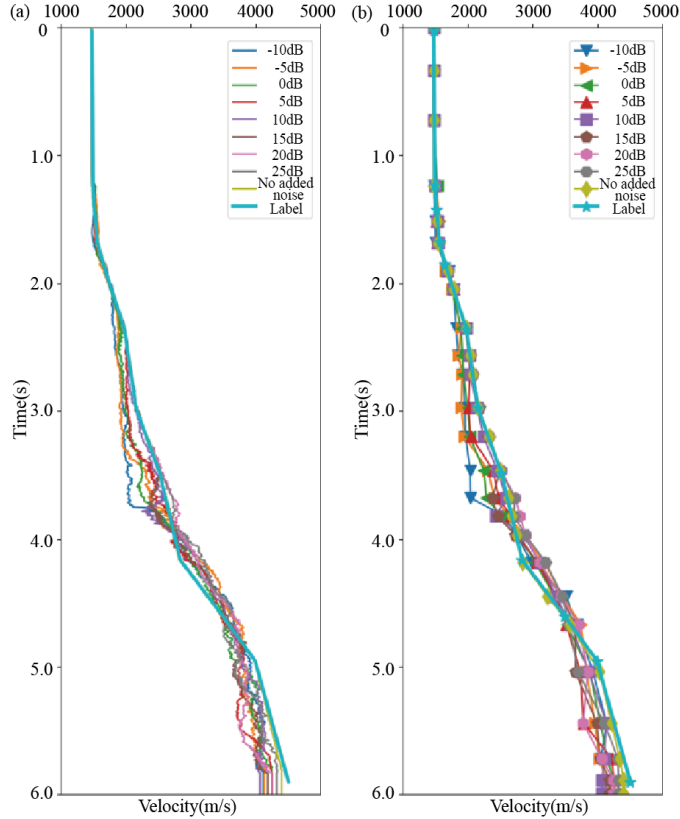

Fig. 9. NMO velocity curves of the predictions in CMP 1179 on field data with different SNR. (a) auto-picks without HLEDS. (b) auto-picks with HLEDS.

### D. Discussion

Compared with labels, the proposed method shows the great potential in automatic velocity picking. In addition, the auto-picks with HLEDS can meet the requirement of picking accuracy in most cases though this strategy will lead to some bias when set of thin horizontal layers exist. We list the difference on detailed time cost for velocity picking with/without HLEDS in Table. I. The time cost of model prediction process is greatly reduced under the HLEDS. In summary, the horizontal layer constrained attention neural network makes automatic semblance velocity picking more efficient and can be well extended to the field data which is short of labels. The HLEDS can be used according to the characteristic of field data and the requirements of efficiency and accuracy in practice.

## IV. CONCLUSIONS

In this paper, we propose an attention neural network combined with a point-to-point regression strategy to achieve automatic semblance velocity picking. In our method, the proposed method can fully utilize the limited labels and the attention neural network can effectively extract essential features hidden in the semblance patch and map it to velocity value. In addition, HLEDS is also incorporated to improve the prediction efficiency. Furthermore, tests on synthetic and field datasets demonstrate that the proposed method can produce reasonable results and maintain better global trend of velocity consistent with labels. Meanwhile, the HLEDS is efficient in prediction which is significant for application in practice. Finally, tests on added Gaussian noise also prove that our method is robust against noise. Actually, the HLEDS also indicates the importance of spatial information hidden in seismic data. one future work is to combine the coherent feature with higher dimension network to optimize the automatic velocity picking strategy.

## V. ACKNOWLEDGMENTS

This work was supported by the National Natural Science Foundation of China under Grant 41974122. We appreciate the open-source software Seismic Unix (SU) which is developed by the Center for Wave Phenomena (http://www.cwp.mines.edu/cwpcodes/). The Hess model is donated to public domain by Faqi Liu and Scott Morton (https://wiki.seg.org/wiki/Hess_VTI_migration_benchmark).

## REFERENCES

[1] M. Taner and F. Koehler, "Velocity spectra-digital computer derivation and applications of velocity functions," Geophysics, vol. 34, no. 6, pp. 859-881, 1969.
[2] J. Huang, J. Cao, G. Chen, and Y. Zhang, "Automate seismic velocity model building through machine learning," SEG Technical Program Expanded Abstracts, pp. 1556-1560, 2020.
[3] C. Fish and T. Kusuma, "A neural network approach to automate velocity picking," SEG Technical Program Expanded Abstracts, pp. 185-188, 1994.
[4] S., Wei, O., Yonglin, Z., Qingcai, H., Jiaqiang, and S., Yaying, "Unsupervised machine learning: K-means clustering velocity semblance autopicking," 80th Annual International Conference and Exhibition, EAGE, Extended Abstracts, pp. 1-5, 2018.
[5] U. Waheed, A. Saleh and M. Sherif, "Machine learning algorithms for automatic velocity picking: K-means vs. DBSCAN," SEG Technical Program Expanded Abstracts, pp. 5110-5114, 2019.
[6] D. Wang, S. Yuan, H. Yuan, H. Zeng and S. Wang, "Intelligent velocity picking based on unsupervised clustering with the adaptive threshold constraint," Chinse J. Geophys. (in Chinese), vol. 64, no. 3, pp.1048-1060, 2021, doi:10.6038/cjg2021O0305.
[7] Geert, Litjens, Thijs, Kooi, Babak, Ehteshami, Bejnordi, Arnaud, Arindra and Adiyoso, "A survey on deep learning in medical image analysis," Medical image analysis, vol. 42, pp. 60-88, 2017.
[8] Z. Zhao, P. Zheng, S. Xu and X. Wu, "Object Detection With Deep Learning: A Review," in IEEE Transactions on Neural Networks and Learning Systems, vol. 30, no. 11, pp. 3212-3232, Nov. 2019, doi: 10.1109/TNNLS.2018.2876865.
[9] M. Park and D. Mauricio, "Automatic velocity analysis using convolutional neural network and transfer learning," Geophysics, vol. 85, no. 1, pp. V33-V43, 2020.
[10] W. Wang, George A. Mcmechan, J. Ma and F Xie. "Automatic velocity picking from semblances with a new deep learning regression strategy: comparison with a classification approach," Geophysics, vol. 86, no. 2, pp. 1-56, 2020.
[11] H. Zhang, P. Zhu, Y. Gu and X. Li, "Automatic velocity picking based on deep learning," SEG Technical Program Expanded Abstracts, pp. 2604-2608, 2019.
[12] M. Araya-Polo, J. Jennings, A. Adler, and T. Dahlke, "Deep-learning tomography," Leading Edge, vol. 37, no. 1, pp. 58–66, 2018. doi: 10.1190/tle37010058.1.

[13] H. Ma, "A Velocity Spectrum Picking Method Based on Detection Fine Tuning Depth Recognition Technology," 2021 IEEE 4th International Conference on Computer and Communication Engineering Technology (CCET), pp. 69-74, 2021, doi: 10.1109/CCET52649.2021.9544220.

[14] A. AlAli and F. Anifowose, "Seismic velocity modeling in the digital transformation era: a review of the role of machine learning," J Petrol Explor Prod Technol, vol. 12, pp. 21-34, 2022, https://doi.org/10.1007/s13202-021-01304-0

[15] X. Wang, R. Girshick, A. Gupta and K. He, "Non-local neural networks," In Proceedings of the IEEE conference on computer vision and pattern recognition, pp. 7794-7803, 2018.

[16] K. He, X. Zhang, S. Ren, and J. Sun, "Deep residual learning for image recognition," in Proc. CVPR, pp.770-778, Jun. 2016.

[17] C. Qiu, B. Wu, D. Meng, X. Zhu, M. Li and N. Qin, "Attention Neural Network Semblance Velocity Auto Picking with Reference Velocity Curve Data Augmentation," 2021 IEEE International Geoscience and Remote Sensing Symposium IGARSS, pp. 4596-4599, 2021, doi: 10.1109/IGARSS47720.2021.9553655.

[18] D. Kingma and J. Ba, "Adam: A method for stochastic optimization," 2014, [Online], Available: https://arxiv.org/abs/1412.6980